\documentstyle[12pt] {article}
\def\lsim{\raise0.3ex\hbox{$<$\kern-0.75em\raise-1.1ex\hbox{$\sim$}}}
\def\gsim{\raise0.3ex\hbox{$>$\kern-0.75em\raise-1.1ex\hbox{$\sim$}}}

\begin {document}

\begin{center}
{\Large {\bf BARYON NUMBER TRANSFER}} \\
\vskip 0.3cm
{\Large {\bf IN HADRONIC INTERACTIONS}} \\

\vskip 1.5 truecm
{\bf G. H. Arakelyan$^1$, A. Capella, A. Kaidalov$^2$ and Yu. M.
Shabelski$^3$}\\
\vskip 0.5 truecm
   Laboratoire de Physique Th\'eorique\footnote{Unit\'e Mixte de Recherche -
CNRS - UMR n$^{\circ}$ 8627},\\ Universit\'e de Paris Sud, B\^atiment 210,
91405 Orsay Cedex, France\\   E-mail: Alphonse.Capella@th.u-psud.fr
\end{center}
\vskip 1.5 truecm
\begin{center}
{\bf ABSTRACT}
\end{center}

The process of baryon number transfer due to string junction propagation
in rapidity is considered. It has a significant effect
in the net baryon production in $pp$ collisions at mid-rapidities and
an even larger effect in the forward hemisphere in the cases
of $\pi p$ and $\gamma p$ interactions. The results of numerical
calculations in the framework of the Quark-Gluon String model are in 
reasonable agreement with the data.

\vskip 1.5 truecm

\vskip 0.3 truecm
\noindent $^1$Permanent address: Yerewan Physics Institute, Armenia and
JINR, Dubna, Russia
E-mail: argev@jerewan1.yerphi.am
\vskip 0.3 truecm
\noindent $^2$Permanent address: Institute of Theoretical and Experimental
Physics, Moscow, Russia \\
E-mail: kaidalov@vitep5.itep.ru
\vskip 0.3 truecm
\noindent $^3$Permanent address: Petersburg Nuclear Physics Institute,
Gatchina, St.Petersburg, Russia
E-mail: shabelsk@thd.pnpi.spb.ru

\vskip 1cm
\noindent LPT-Orsay 01-24 \par
\noindent March 2001
\newpage
\pagestyle{plain}
\noindent{\bf 1. INTRODUCTION}
\vskip 0.5 truecm

The Quark--Gluon String Model (QGSM) and the Dual Parton Model (DPM) are
based on
the Dual Topological Unitarization (DTU) and describe quite reasonably many
features of high energy production processes, including the inclusive
spectra of
different secondary hadrons, their multiplicities, KNO--distributions,
etc., both
in hadron--nucleon and hadron--nucleus collisions \cite{KTM,2r,KTMS,Sh}.
High energy interactions are considered as proceeding via the exchange
of one or several pomerons and all elastic and inelastic processes
result from cutting through or between pomerons \cite{AGK}. The
possibility of exchanging a different number of pomerons introduces
absorptive corrections to the cross sections which are in
agreement with the experimental data on production of hadrons consisting
of light quarks. Inclusive spectra of hadrons are related to the
corresponding fragmentation functions of quarks and diquarks, which
are constructed using the reggeon counting rules \cite{Kai}.

In the present paper we discuss the processes connected with the transfer
of baryon charge over long rapidity distances. In the string models
baryons are considered as configurations consisting of three strings
attached to three valence quarks and connected in a point
called ``string junction" \cite{IOT,RV}. Such a configuration corresponds
in QCD to the following gauge-invariant operator
\begin{equation}
\psi_i(x_1) \psi_j(x_2) \psi_k(x_3) G[P(x_1,x)]^i_{i'}
G[P(x_2,x)]^j_{j'} G[P(x_3,x)]^k_{k'} \varepsilon^{i'j'k'} \; ,
\end{equation}
where
\begin{equation}
G[P(x_1,x)]^i_{j} = \left [ T \exp (g \int_{P(x_1,x)}
A_{\mu} (x) dx^{\mu}) \right ] ^i_j   \; .
\end{equation}

It is very important to understand the role of the string-junction in
the dynamics of high-energy hadronic interactions. Now we have several
different experimental results concerning such processes. First of all
the data \cite{Bren} clearly show that in the forward hemisphere the
number of secondary protons produced in $\pi^+p$ interactions is
significantly larger than the number of $\bar{p}$ produced in $\pi^-p$
collisions. This difference can not be described \cite{Sh} without the
assumption that the baryon charge is transferred from the target proton to
the pion hemisphere.

Similar data on the differences of $p - \bar{p}$ yields in
$\frac12(\pi^+p + \pi^-p)$ collisions at 158 GeV/c were presented by
the NA49 Coll. \cite{NA49}.

A second group of data concerns the energy dependence of the differences
in yields of the protons and antiprotons at 90$^o$ (i.e. at $x_F$ = 0) at
ISR energies \cite{ISR}.

Another sample of data includes the measurements of hyperon production
asymmetries in 500 GeV/c $\pi^-$-nucleus interactions \cite{ait1}.

Finally, the proton-antiproton asymmetry in photoproduction was recently
measured at HERA \cite{H1}.

In this paper we present a simultaneous description of all these data and
extract information on the properties of the string junction dynamics.

\vskip 0.9 truecm \noindent{\bf 2. INCLUSIVE SPECTRA OF SECONDARY HADRONS IN
QGSM} \vskip 0.5 truecm

As mentioned above high energy hadron--nucleon and hadron--nucleus interactions
are considered in the QGSM and in DPM as proceeding via the exchange of one or
several pomerons. Each pomeron corresponds to a cylindrical diagram 
(see Fig. 1a),
and thus, when cutting a pomeron two showers of secondaries are 
produced (Fig. 1b). The
inclusive spectrum of secondaries is determined by the convolution of diquark,
valence and sea quark distributions $u(x,n)$ in the incident particles and the
fragmentation functions $G(z)$ of quarks and diquarks into secondary hadrons.
The diquark and quark distribution functions depend on the number $n$ of cut
pomerons in the considered diagram. In the following we use the formalism of
QGSM. In the case of a nucleon target the inclusive spectrum of a secondary
hadron $h$ has the form \cite{KTM}: \begin{equation} \frac{x}{\sigma_{inel}}
\frac{d\sigma}{dx} =\sum_{n=1}^{\infty}w_{n}\phi_{n}^{h}(x)\ \ , \end{equation}
where the functions $\phi_{n}^{h}(x)$ determine the contribution of
diagrams with $n$ cut pomerons and $w_{n}$ is the probability of this
process. Here we neglect the contributions of diffraction dissociation
processes which are comparatively small in most of the processes
considered below. It can be accounted for separately \cite{KTM,2r,Sh}.

For $pp$ collisions
\begin{equation}
\phi_{pp}^{h}(x) = f_{qq}^{h}(x_{+},n)f_{q}^{h}(x_{-},n) +
f_{q}^{h}(x_{+},n)f_{qq}^{h}(x_{-},n) +
2(n-1)f_{s}^{h}(x_{+},n)f_{s}^{h}(x_{-},n)\ \  ,
\end{equation}
\begin{equation}
x_{\pm} = \frac{1}{2}[\sqrt{4m_{T}^{2}/s+x^{2}}\pm{x}]\ \ ,
\end{equation}
where $f_{qq}$, $f_{q}$ and $f_{s}$ correspond to the contributions of
diquarks, valence and sea quarks respectively. They are determined by
the convolution of the diquark and quark distributions with the
fragmentation functions, e.g.,
\begin{equation}
f_{q}^{h}(x_{+},n) = \int_{x_{+}}^{1} u_{q}(x_{1},n)G_{q}^{h}(x_{+}/x_{1})
dx_{1}\ \ .
\end{equation}
In the case of a meson beam the diquark contributions in Eq. (4) should
be changed by the contribution of valence antiquarks:
\begin{equation}
\phi_{\pi p}^{h}(x) = f_{\bar{q}}^{h}(x_{+},n)f_{q}^{h}(x_{-},n) +
f_{q}^{h}(x_{+},n)f_{qq}^{h}(x_{-},n) +
2(n-1)f_{s}^{h}(x_{+},n)f_{s}^{h}(x_{-},n)\ \  .
\end{equation}

The diquark and quark distributions as well as the fragmentation
functions are determined from Regge intercepts. Their expressions are given
in Appendix 1. \par

The net baryon charge can be obtained from the fragmentation of the diquark
giving rise to a leading baryon (Fig. 2a). A second possibility is to
produce a
(leading) meson in the first break-up of the string and the baryon in a
subsequent break-up (Fig.~2b). \par

   As discussed above, in the approach \cite{IOT,RV} the baryon consists of
three
valence quarks together with string junction (SJ), which is conserved
during the
interaction\footnote{At very high energies one or even several SJ pairs
can be
produced.}.

This gives a third possibility for secondary net baryon production in
non-diffractive hadron-nucleon interactions (Fig. 2c). \par

The secondary baryon consists of the SJ together with two
valence and one sea quarks (Fig.~2a), one valence and two sea quarks
(Fig.~2b) or three sea quarks (Fig.~2c). The fraction of the incident 
baryon energy
carried by the secondary baryon decreases from a) to c), whereas the mean
rapidity gap between the incident and secondary baryon increases.

The probability to find a comparatively slow SJ in the case of Fig. 2c
can be estimated from the data on $\bar{p}p$ annihilation into
mesons (see Figs. 1c, d). This probability is known experimentally only at
comparatively small energies where it is proportional to
$s^{\alpha_{SJ}-1}$ with $\alpha_{SJ} \sim 0.5$.

However, it has been argued \cite{14r} that the annihilation cross section
contains a small piece which is independent of $s$ and thus $\alpha_{SJ}
\sim
1$. Irrespectively of the value of $\alpha_{SJ}$, the contribution of the
graph
in Fig.~2c corresponds to annihilation and, thus, has a small coefficient
which
will be denoted by $\varepsilon$. In our calculation we shall use

\begin{equation}
\alpha_{SJ} = 0.5
\end{equation}

\noindent and will treat $\varepsilon$ as a free parameter. A contribution
with
$\alpha_{SJ} \sim 1$, if present at all, is expected to be important only
at
very high energies (see a discussion of this point in section 5). The values of
$\alpha_{SJ}$ and $\varepsilon$ can only be determined with the help 
of accurate data.

\vskip 0.9 truecm
\noindent{\bf 3. COMPARISON WITH THE DATA}
\vskip 0.5 truecm

The mechanism of the baryon charge transfer via SJ without valence 
quarks (Fig.~2c) was not accounted for in
previous papers \cite{KTM,2r,KTMS,Sh}.

The data at comparatively low energies ($\sqrt{s} \sim 15 \div 40$~GeV)
can be
described with $\alpha_{SJ} = 0.5$. Unfortunately, the value of
$\varepsilon$
cannot be determined in a unique way. Many sets of data can be described
with
$\varepsilon = 0.05$. However, other data sets favor a value four times
larger,
$\varepsilon = 0.2$. Because of that, we will present our results for these
two
values of $\varepsilon$, leaving all other parameters in the model
unchanged.
Thus, the results for meson and antibaryon production are the same in the
two
cases.

The inclusive spectra of secondary protons and antiprotons produced
in $pp$ collisions at lab. energies 100 and 175 GeV \cite{Bren} are
shown in Figs. 3a and 3b together with the curves calculated in the
QGSM. The agreement of our results with $\varepsilon = 0.05$ with the
data is quite reasonable, on the same level (or even better) as in the
previous papers which did not incorporate the SJ  mechanism of 
Fig.~2c. The variant with $\varepsilon = 0.2$ gives too large
multiplicity of secondary protons at small $x_F$. The description of
the antiproton yields is reasonable.

The data of Ref. \cite{AB} are in some regions of $x_F$ in disagreement with
the data of Ref. \cite{Bren} as well as with our calculations. However, one
can see from Figs. 3c and 3d that the experimental points can not be
described by any smooth curve. Probably this is connected with the use of
different detectors for different $x_F$ regions and to the influence of the
trigger in \cite{AB}. Again, the value $\varepsilon = 0.05$ is
preferable.

The data on secondary proton and antiproton production in $pp$
collisions at ISR energies \cite{ISR} at $90^o$ in c.m.s. are presented
in Figs. 3e and 3f. Their differences, which are more sensitive to the
baryon charge transfer, are presented in Fig. 3g.
One can see that the last data, as well as the yields of protons and
antiprotons separately, are described quite reasonably by QGSM with
$\varepsilon =
0.2$.  However, it is necessary to note, that the systematic errors in
\cite{ISR} are of the order of 30~\%, so the value
$\varepsilon = 0.05$ can not be excluded. Thus the disagreement 
between ISR data
\cite{ISR} and more recent data \cite{Bren,AB} on spectra of protons 
in $pp$-collisions
does not allow to determine uniquely the value of $\varepsilon$.

The data on baryon production in the pion fragmentation region \cite{Bren}
are presented in Fig. 4. The spectra of antiprotons produced in $\pi^-p$
collisions, shown in Fig. 4a, allows one to fix the fragmentation
function of a quark into baryon/antibaryon. If the contribution of the
baryon charge transfer were negligibly small, the inclusive spectra of
reactions $\pi^-p \to \bar{p}X$ and $\pi^+p \to pX$ in the pion
fragmentation region would be practically the same \cite{Sh}. Actually,
the data for the second reaction are significantly higher than for the
first one providing evidence for the baryon charge transfer due to the
SJ propagation. The difference of the inclusive spectra in the two considered
processes allows one to estimate quantitatively the contribution of
the baryon charge transfer, and the parametrization of these processes
given in Appendix leads to a reasonable description of proton yields
in $\pi^+p$ collisions with $\varepsilon = 0.2$. The value
$\varepsilon = 0.05$ seems to be too small here. Note, however, that our
factorized formulae, Eqs. 4 and 7, imply that the slowed down proton is
made out of SJ and three sea quarks (see Fig.~2c). Thus, they should be modified
for the reaction $\pi^+p \to pX$ in the pion fragmentation region due to the
possibility of SJ recombination with a pion valence quark -- which would change the
ratio of protons and neutrons. A simple estimate based on the quark combinatorics given
in Appendix 1 leads to an increase of the proton yield in the reaction 
$\pi^+p \to pX$ by about 50~\%. In this case the value of $\varepsilon = 0.05$ can be
consistent with the data. This problem disappears if we consider the sum of
$\pi^+p$ and $\pi^-p$. Preliminary data on ${1 \over 2} (\pi^+p + \pi^-p) \to
(p-\bar{p})X$ were obtained by the NA49 Coll. \cite{NA49}. One can 
see from Fig.4c, that they are in good agreement with our calculations with
$\varepsilon = 0.05$ and in total disagreement with $\varepsilon = 0.2$.


The data on $\Lambda$ and $\bar{\Lambda}$ production in $pp$ collisions
\cite{Ch,Bo,She,Ki}, presented in Fig. 5, are also in agreement with
QGSM  and
the value $\varepsilon = 0.05$ is again favored.

In Fig.~6 we show the data \cite{ait1} on the asymmetry of strange baryons
produced in $\pi^-$ interactions\footnote{These data were obtained
from pion interactions on a nuclear target where different materials were
used in a very complicated geometry. We assume that the nuclear effects
are small in the asymmetry ratio (9), and compare the pion-nucleus
data with calculations for $\pi^-p$ collisions.} at 500 GeV/c. The
asymmetry is determined as
\begin{equation}
A(B/\bar{B}) = \frac{N_B - N_{\bar{B}}}{N_B + N_{\bar{B}}}
\end{equation}
for each $x_F$ bin.

The theoretical curves for the data on all asymmetries calculated with
$\varepsilon = 0.05$ are in reasonable agreement with the data with the
exception of $\Xi^-/\Xi^+$ at positive $x_F$. However in this region the effect
of recombination of SJ with valence (d) quark of $\pi^-$ can be important and
can lead to an increase of cross section analogous to the one in $\pi^+p \to
pX$ discussed above. In the case of $\Omega/\bar{\Omega}$ production 
we predict a non-zero
asymmetry in agreement with experimental data. Let us note that the last
asymmetry is absent, say, in the naive quark model because $\Omega$ and
$\bar{\Omega}$ have no common valence quarks with the incident particles. The
value $\varepsilon = 0.2$ seems to be excluded.

Preliminary data on $p/\bar{p}$ asymmetry in $ep$ collisions at HERA were
presented by the H1 Collaboration \cite{H1}. Here the asymmetry is defined as

\begin{equation}
A_B = 2 \frac{N_p - N_{\bar{p}}}{N_p + N_{\bar{p}}} \; ,
\end{equation}
i.e. with an additional factor 2 in comparison with Eq. (9). The
experimental value of $A_B$ is equal to $8.0 \pm 1.0 \pm 2.5$ \% \cite{H1} for
secondary baryons produced at $x_F \sim 0.04$ in the $\gamma p$ c.m. 
frame. QGSM
with $\varepsilon = 0.05$ predicts here 2.9~\%, i.e. a smaller value, 
whereas the calculation with $\varepsilon = 0.2$ gives
the value 7.7~\%, in agreement with the
data.

\vskip 0.9 truecm
\noindent{\bf 5. CONCLUSIONS}
\vskip 0.5 truecm

We have demonstrated that experimental data on high-energy hadronic
interactions support the possibility of baryon charge transfer over large
rapidity distances. Probably, the most important are the baryon-antibaryon
asymmetry in $\Omega$ and $\bar{\Omega}$ \cite{ait1} in $\pi^-p$ collisions,
where both secondary particles have no common valence quarks with the incident
particles. This asymmetry is provided by baryon charge transfer due to
string junction diffusion.

Also the production of net baryons in $\pi p$ and $\gamma p$ interactions
in the
projectile hemisphere provides good evidence for such a mechanism. \par

As for the values of the parameters $\alpha_{SJ}$ and $\varepsilon$ which
govern
the baryon charge transfer we have seen that the data, at comparative low
energies ($\sqrt{s} \sim 15 \div 40$~GeV), without providing a clear cut
answer,
strongly favor the values $\alpha_{SJ} = 0.5$ and $\varepsilon = 0.05$.
Indeed,
as discussed above, the data which favor $\varepsilon = 0.2$ have either a
large
systematic error \cite{ISR} or correspond to situations where effects, not
incorporated in the present version of the model, are expected to be
important
(as in $\pi^+p \to pX$~; see the discussion in Section 3).

At HERA energies the situation is different. Here the observed
asymmetry is better described with $\varepsilon = 0.2$. Note, 
however, that HERA
data are
preliminary and have rather large errors (the prediction for asymmetry with
$\varepsilon = 0.05$ deviates from the data by about two standard deviations).
If this discrepancy is real can it be solved~? The data at $\sqrt{s}
\ \lsim$ 40~GeV that we have examined provide information for the transfer of
baryon charge over about five rapidity units. In contrast, the HERA 
data provide
information for the corresponding transfer over more than seven 
units. In ref. \cite{BKP} HERA data were described under the 
assumption that there is a
component with $\alpha_{SJ} \approx 1$ and a very small strength. Another
possibility would be to use $\alpha_{SJ} \simeq 0.5$ and to assume,
following
Ref. \cite{22r}-\cite{CS}  that the strength of the $SJ$ transfer
contribution increases with the number of inelastic collisions. In this way the
asymmetry would increase with energy since the average number of inelastic
collisions, determined from Eqs. (42) and (43), increases with $s$. (This
mechanism also allows to explain the increase of stopping in heavy ion
collisions with increasing centrality) \cite{22r}-\cite{26r}.

In conclusion, the study of the transfer of baryon charge over large rapidity
distances is very important. Good experimental data in $pp$, $pA$ and $AB$
collisions for different centralities are needed in order to understand the
dynamics of this mechanism.\\

Acknowledgments: We thank H.G.Fisher from NA49 Collaboratiomn for providing 
the preliminary data in Fig. 4c prior to publication.

This paper was supported by grant NATO OUTR. LG 971390 and NATO PSTCLG 977275.

\vskip 0.9 truecm \noindent{\bf APPENDIX. QUARK AND DIQUARK DISTRIBUTIONS AND
THEIR FRAGMENTATION FUNCTIONS} \vskip 0.5 truecm

In the present calculations we use quark and diquark distributions in
the proton for one-pomeron exchange given by the correspondent Regge
behaviour \cite{KTM} 
\begin{eqnarray} 
u_{uu}(x,n) &=& C_{uu}x^{\alpha_R-2\alpha_B+1 }(1-x)^{-\alpha_R}\;, \\ 
u_{ud}(x,n) &=& C_{ud}x^{\alpha_R - 2 \alpha_B }(1-x)^{-\alpha_R}\;, \\ 
u_{u}(x,n) &=& C_{u}x^{-\alpha_R}(1-x)^{\alpha_R - 2\alpha_B +1}\;, \\ 
u_{d}(x,n) &=& C_{d}x^{-\alpha_R}(1-x)^{n+\alpha_R-2\alpha_B}\;, \\
\end{eqnarray}

In the case of multipomeron exchange the distributions of valence quarks
and diquarks are softened due to the appearance of sea quark contribution.
Aslo it is necessary to account that $d$-quark distribution is more
soft in comparison with $u$-quark one. There is some freedom \cite{KTMS} 
how to account for these effects. We use the simplest way and write
diquark and quark distributions (for $\alpha_R = 0.5$ and 
$\alpha_B = -0.5$) as 
\begin{eqnarray}
u_{uu}(x,n) &=& C_{uu}x^{\alpha_R-2\alpha_B+1
}(1-x)^{frac43(n-1)-\alpha_R}\;, \\ 
u_{ud}(x,n) &=& C_{ud}x^{\alpha_R-2\alpha_B }(1-x)^{n-1-\alpha_R}\;, \\
\\ u_{u}(x,n) &=& C_{u}x^{-\alpha_R}(1-x)^{n+\alpha_R-2\alpha_B+1}\;, \\
\\ u_{d}(x,n) &=& C_{d}x^{-\alpha_R}
(1-x)^{\frac43(n-1)+\alpha_R-2\alpha_B+1}\;, \\
\\u_{\overline{u}}(x,n) &=& u_{\overline{d}}(x,n)= C_{\overline{u}}
x^{-\alpha_R} \nonumber
\\&\times &[(1-x)^{n+\alpha_R -2\alpha_B-1} -
\delta /2(1-x)^{n+2\alpha_R-2\alpha_B-1}] \; , \; n>1\ \ ,\\
u_{s}(x,n) &=& C_{s}x^{-\alpha_R}(1-x)^{n+2\alpha_R-2\alpha_B-1} \; , \;
n>1\ \ .
\end{eqnarray}

In the case of a pion beam we use
\begin{equation}
u_{q}(x,n) = C_{q}x^{-\alpha_R}(1-x)^{n-\alpha_R -1}\ \ , \\
u_{\bar{q}}(x,n) = C_{\bar{q}}x^{-\alpha_R}(1-x)^{n-\alpha_R-1}\ \ ,\\
\end{equation}
for valence quarks, and
$$u(x,n) =  Cx^{-\alpha_R}(1-x)^{n-\alpha_R -1}[1-\delta \sqrt{1-x}]
   \; , \; n>1 \; ,$$
\begin{equation}
u_{s}(x,n) = C_{s}x^{-\alpha_R}(1-x)^{n-1} \; , \; n>1\ \ ,
\end{equation}
for sea quarks,
where $\delta =0.2$ is the relative probability to find a strange quark in
the
sea. The values of $\alpha_R$ and $\alpha_B$ are given in Table 1. The
factors
$C_{i}$ are determined from the normalization condition 
\begin{equation}
\int_{0}^{1} u_{i}(x,n)dx = 1\ \ .
\end{equation}
and sum rule
\begin{equation}
\int_{0}^{1}\sum_i u_{i}(x,n)xdx = 1\ \ .
\end{equation}
is fulfilled.

The fragmentation functions of quarks and diquarks were changed a little
in compari\-son with Refs. \cite{KTM,Sh} to obtain a better agreement
with the existing experimental data, because the old functions 
\cite{KTM,Sh} correspond to $\varepsilon$ = 0. We use the quark
fragmentation functions in the form
\begin{equation}
G_{u}^{p} = G_{d}^{p} =
a_{\bar{N}}(1-z)^{\lambda + \alpha_R - 2 \alpha_B}(1+a_{1}z^{2}) \;\; ,
G_{u}^{\bar{p}} = G_{d}^{\bar{p}} = (1-z) G_{u}^{p}
\end{equation}

\begin{equation}
G_{u}^{\Lambda} = G_d^{\Lambda} = \frac{a_{\bar{\Lambda}}}{a_{\bar{N}}}
(1-z)^{\Delta \alpha} G_u^p \;\; ,
G_u^{\bar{\Lambda}} = (1-z) G_d^{\Lambda}
\end{equation}

\begin{equation}
G_d^{\Xi^-} = \frac{a_{\bar{\Xi}}}{a_{\bar{\Lambda}}}
(1-z)^{\Delta \alpha} G_u^p \;\; ,
G_u^{\Xi^-} = G_u^{\bar{\Xi}} = (1-z) G_d^{\Xi^-}
\end{equation}

\begin{equation}
G_{u}^{\Omega} = G_d^{\Omega} = G_u^{\bar{\Omega}} = G_d^{\bar{\Omega}}
= \frac{a_{\bar{\Omega}}}{a_{\bar{\Xi}}}
(1-z)^{\Delta \alpha} G_u^{\Xi} \;\; ,
\end{equation}

with
\begin{equation}
\Delta \alpha = \alpha_{\rho} - \alpha_{\phi} = 1/2 \;\; ,~~~~               
\lambda=2\alpha^{\prime} < p_{t}^2>=0.5 \; .
\end{equation}

Diquark fragmentation functions have more complicate forms. They contain
two contributions. The first one corresponds to the central
production of
a $B\bar{B}$ pair and can be described by the previous formulas. They have
the form:

\begin{equation}
G_{uu}^p = G_{ud}^p = G_{uu}^{\bar{p}} = G_{ud}^{\bar{p}}
= a_{\bar{N}}(1-z)^{\lambda-\alpha_R + 4(1-\alpha_B)} \ \ ,
\end{equation}

\begin{equation}
G_{uu}^{\Lambda} = G_{ud}^{\Lambda} = G_{uu}^{\bar{\Lambda}}
= G_{ud}^{\bar{\Lambda}} = \frac{a_{\bar{\Lambda}}}{a_{\bar{N}}}
(1-z)^{\Delta \alpha} G_{uu}^p \;\; ,
\end{equation}

\begin{equation}
G_{uu}^{\Xi^-} = G_{ud}^{\Xi^-} = G_{uu}^{\bar{\Xi}} = G_{ud}^{\bar{\Xi}}
= \frac{a_{\bar{\Xi}}}{a_{\bar{\Lambda}}} (1-z)^{\Delta \alpha}
G_{uu}^{\Lambda} \;\; ,
\end{equation}

\begin{equation}
G_{uu}^{\Omega} = G_{ud}^{\Omega} = G_{uu}^{\bar{\Omega}}
= G_{ud}^{\bar{\Omega}}
= \frac{a_{\bar{\Omega}}}{a_{\bar{\Xi}}} (1-z)^{\Delta \alpha}
G_{uu}^{\Xi}
\end{equation}
with the same $\Delta \alpha$ Eq. (29).

The second contribution is connected with the direct fragmentation of the
initial baryon into the secondary one with conservation of the string
junction. As discussed above, there exists three
different types of such contributions (Figs.~2a-2c). Obviously,
in the case of $\Xi$ production only two possibilities exist with string
junction
plus either one valence quark and two sea quarks or three sea quarks.
In the case
of production of a secondary baryon having no common quarks with the incident
nucleons only the bare string junction without valence quarks can
contribute (Fig.~2c).

All these contributions are determined by Eqs. similar to Eq. (6)
with the corresponding fragmentation functions given by
\begin{equation}
G_{uu}^p = G_{ud}^p = a_N \sqrt{z} [v_0\varepsilon (1-z)^2 +
v_q z^{3/2} (1-z) + v_{qq}z^2] \; ,
\end{equation}
\begin{equation}
 G_{ud}^{\Lambda} = a_N \sqrt{z}
[v_0\varepsilon (1-z)^2 + v_q z^{3/2} (1-z) + v_{qq}z^2]
(1-z)^{\Delta \alpha}, G_{uu}^{\Lambda} =(1-z)G_{ud}^{\Lambda} \; 
\end{equation}
\begin{equation}
G_{d,SJ}^{\Xi^-} = a_N \sqrt{z} [v_0\varepsilon (1-z)^2 +
v_q z^{3/2} (1-z)] (1-z)^{2\Delta \alpha}, ~G_{u,SJ}^{\Xi^-}= 
(1-z)G_{d,SJ}^{\Xi^-}\; ,
\end{equation}
\begin{equation}
G_{SJ}^{\Omega} = a_N v_0\varepsilon \sqrt{z}
(1-z)^{2+3\Delta \alpha} \; .
\end{equation}

The factor $\sqrt{z}$ is really $z^{1-\alpha_{SJ}}$ with
$\alpha_{SJ} = 1/2$ (8). As for the factor $\sqrt{z} z^{3/2}$ of the
second term it
is just $2(\alpha_R - \alpha_B)$ \cite{KTM}. For the third term we have
just added
an extra factor $z^{1/2}$.

The probabilities of transition into the secondary baryon of SJ without
valence quarks, $I_3$, SJ plus one valence quark, $I_2$, and
SJ plus a valence diquark, $I_1$, were taken from
the simplest quark combinatorics \cite{CS}. Assuming that the strange
quark suppression
is the same in all these cases, we obtain for the relative yields of different
baryons from SJ fragmentation without valence quarks~:
\begin{equation}
I_3 = 4L^3 : 4L^3 : 12L^2S : 3LS^2 : 3LS^2 : S^3
\end{equation}
for secondary $p$, $n$, $\Lambda + \Sigma$, $\Xi^0$, $\Xi^-$ and
$\Omega$, respectively.

For $I_2$ we obtain
\begin{equation}
I_{2u} = 3L^2 : L^2 : 4LS : S^2 : 0
\end{equation}
and
\begin{equation}
I_{2d} = L^2 : 3L^2 : 4LS : 0 : S^2
\end{equation}
for secondary $p$, $n$, $\Lambda + \Sigma$, $\Xi^0$ and  $\Xi^-$.

For $I_1$ we have
\begin{equation}
I_{1uu} = 2L : 0 : S
\end{equation}
and
\begin{equation}
I_{1ud} = L : L : S
\end{equation}
for secondary $p$, $n$ and $\Lambda + \Sigma$. The ratio $S/L$
determines the strange suppression factor and $2L + S$ = 1.
In the numerical calculations we used $S/L = 0.2$.

In agreement with experimental data we assume that
$\Sigma^+ + \Sigma^- = 0.6\Lambda$ \cite{CS} in Eqs.~(38)-(40). This
is obtained when, say, an $uus$ state can turn into $\Sigma^+$ with
probability 3/4 and into $\Lambda$ (via
$\Sigma^+(1385) \to \Lambda \pi^+$ decay) with probability 1/4
\footnote{These probabilities are in disagreement with simplest quark
statistic rules \cite{AKNS}.}. In the case of  Eqs.~(41) and (42)
$\Sigma^-$ can not be produced and we assume that
$\Sigma^+ = 0.3\Lambda$.

The values of $v_0$, $v_q$ and $v_{qq}$ are determined directly by the
corresponding coefficients of Eqs.~(38)-(42), together with the
probabilities
to fragment a $qqs$ system into $\Sigma^+ + \Sigma^-$ and into $\Lambda$,
given above. For example, we have for incident $uu$ diquark and secondary
proton :
\begin{equation}
v_0 = 4L^3 \;\;, v_q = 3L^2 \;\;,v_{qq} = 2L \; ;
\end{equation}
for incident $ud$ diquark and secondary proton :
\begin{equation}
v_0 = 4L^3 \;\;, v_q = 2L^2 \;\;,v_{qq} = L \; ;
\end{equation}
for incident $uu$ diquark and secondary $\Lambda$ :
\begin{equation}
v_0 = \frac{12}{1.6}S L^2 \;\;, v_q = \frac4{1.6}S L \;\;,
v_{qq} = \frac14 S \; ;
\end{equation}
for incident $ud$ diquark and secondary $\Lambda$ :
\begin{equation}
v_0 = \frac{12}{1.6}S L^2 \;\;, v_q = \frac4{1.6}S L \;\;,
v_{qq} = S \; ;
\end{equation}
for incident $u$ quark and secondary $\Xi^-$ :
\begin{equation}
v_0 = 3 S^2 L \;\;, v_q = 0 \; ;
\end{equation}
for incident $d$ quark and secondary $\Xi^-$ :
\begin{equation}
v_0 = 3 S^2 L \;\;, v_q = S^2 \; ;
\end{equation}
and incident SJ and secondary $\Omega^-$ :
\begin{equation}
v_0 = S^3 \;.
\end{equation}

The probability for a process to have $n$ cutted pomerons was calculated
using the quasi\--ei\-ko\-nal approximation \cite{KTM,TM}:
\begin{equation}
w_{n} = \sigma_{n}/\sum_{n=1}^{\infty}\sigma_{n}  \;  , \;
\sigma_{n} = \frac{\sigma_{P}}{nz} (1-e^{-z}
\sum_{k=0}^{n-1}\frac{z^{k}}{k!
})\ \  ,
\end{equation}
\begin{equation}
z = \frac{2C\gamma}{R^{2}+\alpha^{\prime}\xi}e^{\Delta\xi} \; ,
\; \sigma_{P} = 8\pi
\gamma e^{\Delta\xi} \; , \; \xi = \ln(s/1\ {\rm GeV}^{2})  \ \ ,
\end{equation}
with parameters
\vskip 0.3 truecm
\begin{center}
$\Delta = 0.139 \; , \; \alpha^{\prime} = 0.21\ {\rm GeV}^{-2} \; ,
\; \gamma_{pp} = 1.77\ {\rm GeV}^{-2}
\; , \; \gamma_{\pi p} = 1.07 \ {\rm GeV}^{-2}\; ,$\newline
$R_{pp}^{2} = 3.18\ {\rm GeV}^{-2} \; , \; R_{\pi p}^{2} = 2.48\ {\rm GeV}
^{-2} \;
, \; C_{pp} = 1.5 \; , \; C_{\pi p} = 1.65\ \ .$
\end{center}
\vskip 0.3 truecm

The model parameters for quark and diquark distributions and their
fragmentation are presented in Table 1. They were mainly taken from
the description of the data in Refs. \cite{KTM,Sh}.

\newpage

\begin{center}
{\bf Table 1}
\end{center}
\vspace{15pt}
The values of the parameters used for the calculations in QGSM.
\begin{center}
\vskip 12pt
\begin{tabular}{|c|c|}\hline

Parameter       & value   \\   \hline

$a_N$           & 1.8   \\

$a_{\bar{N}}$   & 0.18  \\

$a_1$           & 12    \\

$\alpha_R $     & 0.5   \\

$\alpha_B $     & -0.5  \\
\hline
\end{tabular}
\end{center}

\vskip 3.0cm

\newpage

\begin{center}
{\bf Figure captions}
\end{center}
\vskip 0.5 truecm

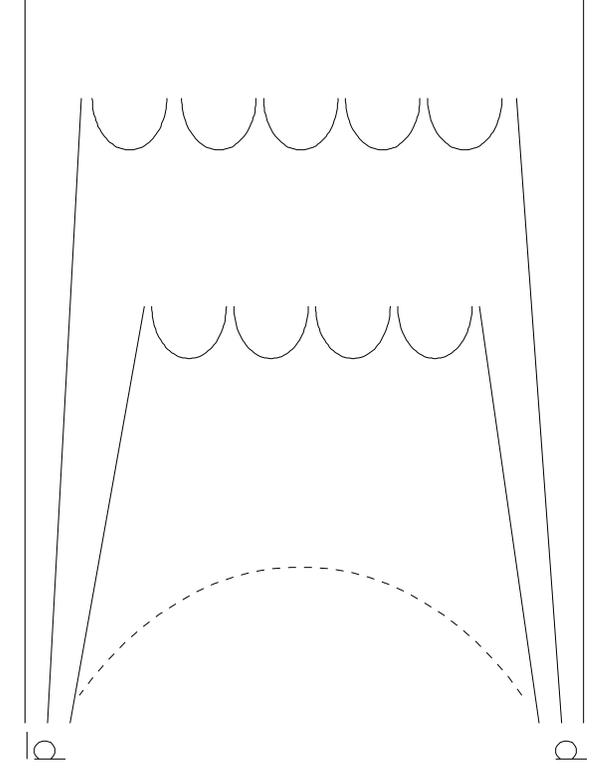
{\bf Fig. 1.} Cylindrical diagram corresponding to the one--pomeron
exchange contri\-bution to elastic $\bar{p}p$ scattering (a) and its cut
which determines the contribution to inelastic $\bar{p}p$ annihilation
cross section (b) (string-junction is indicated by a dashed line). c) the
diagram for elastic $\bar{p}p$-scattering with an exchange by SJ in the
$t$-channel and its $s$-channel discontinuity d).\\

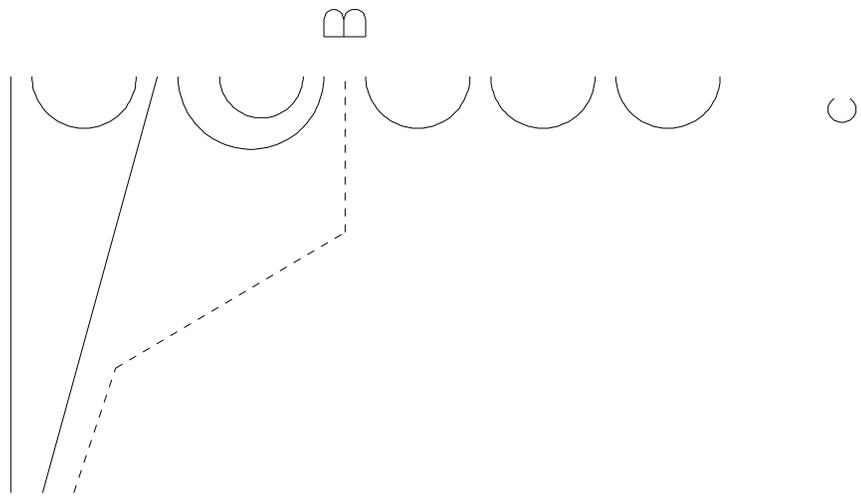
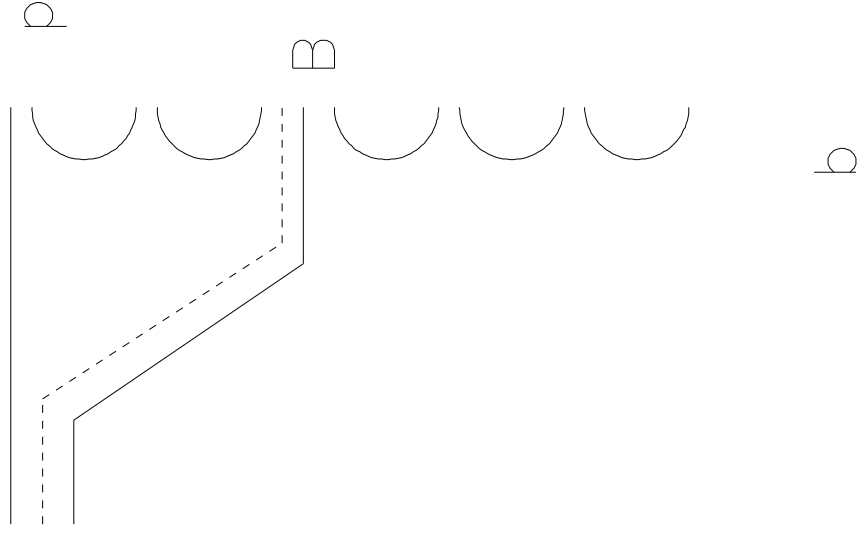
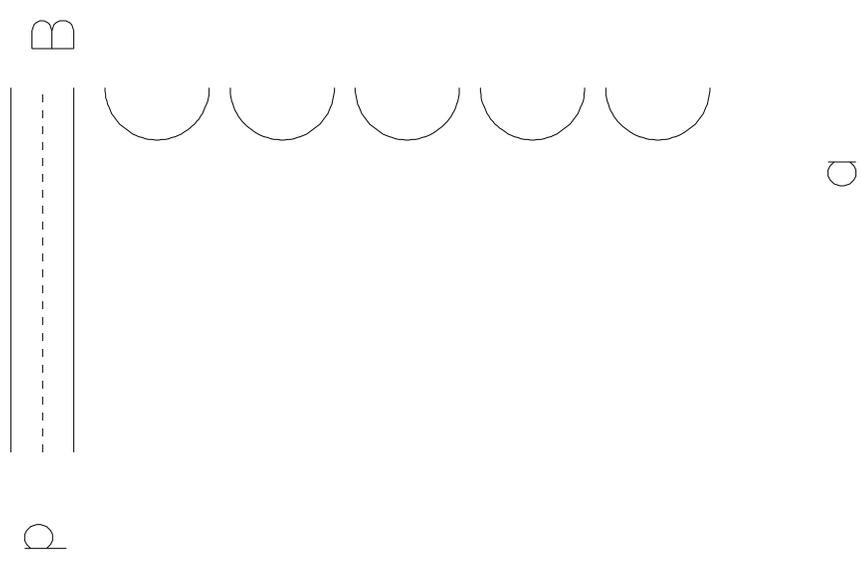
{\bf Fig. 2.} Three different possibilities of secondary baryon
production in $pp$ interac\-tions: string junction together with two
valence and one sea quark (a), together with one valence and two sea quarks
(b), together with three sea quarks (c). \\

{\bf Fig. 3.} The spectra of secondary protons (a), and antiprotons (b) in
$pp$ collisions at 100 and 175 GeV/c \cite{Bren} and at 400 GeV/c
\cite{AB} (c) and (d) and their description by QGSM. Secondary proton
(e) and antiproton (f) yields at ISR energies \cite{ISR} at $90^o$ in
c.m.s. and their difference (g) together with QGSM model predictions.
In all cases the calculations with $\epsilon$ = 0.05 are shown by solid
curves and the variants with $\epsilon$ = 0.2 are shown by dashed
curves. \\

{\bf Fig. 4.} The spectra of secondary antiprotons in $\pi^-p$ collisions
(a) and of protons in $\pi^+p$ collisions (b) at lab. energies 100 and
175 GeV \cite {Bren} and its description by QGSM. The differences $dN/dx$
of $p - \bar{p}$ yields in $\frac12(\pi^+p + \pi^-p)$ collisions at 158
GeV/c compared to preliminary data from NA49 \cite{NA49} (c).
In all cases the calculations with $\epsilon$ = 0.05 are shown by solid curves 
and the variants with $\epsilon$ = 0.2 are shown by dashed curves. \\


{\bf Fig. 5.} The spectra of secondary $\Lambda$ (a) and $\bar{\Lambda}$
(b) in $pp$ collisions (data are taken from \cite{2r}) and its
description by QGSM model. The calculations with $\epsilon$ = 0.05 are
shown by solid curves and a variant with $\epsilon$ = 0.2 are shown by
dashed curves. \\

{\bf Fig. 6.} The asymmetries of secondary $\Lambda/\bar{\Lambda}$ (a),
$\Xi^-/\Xi^+$ (b), and $\Omega/\bar{\Omega}$ (c), in $\pi^-p$
collisions at 500 GeV/c \cite{ait1} and its description by QGSM model.
In all cases the calculations with $\epsilon$ = 0.05 are shown by solid
curves and the variants with $\epsilon$ = 0.2 are shown by dashed
curves. \\

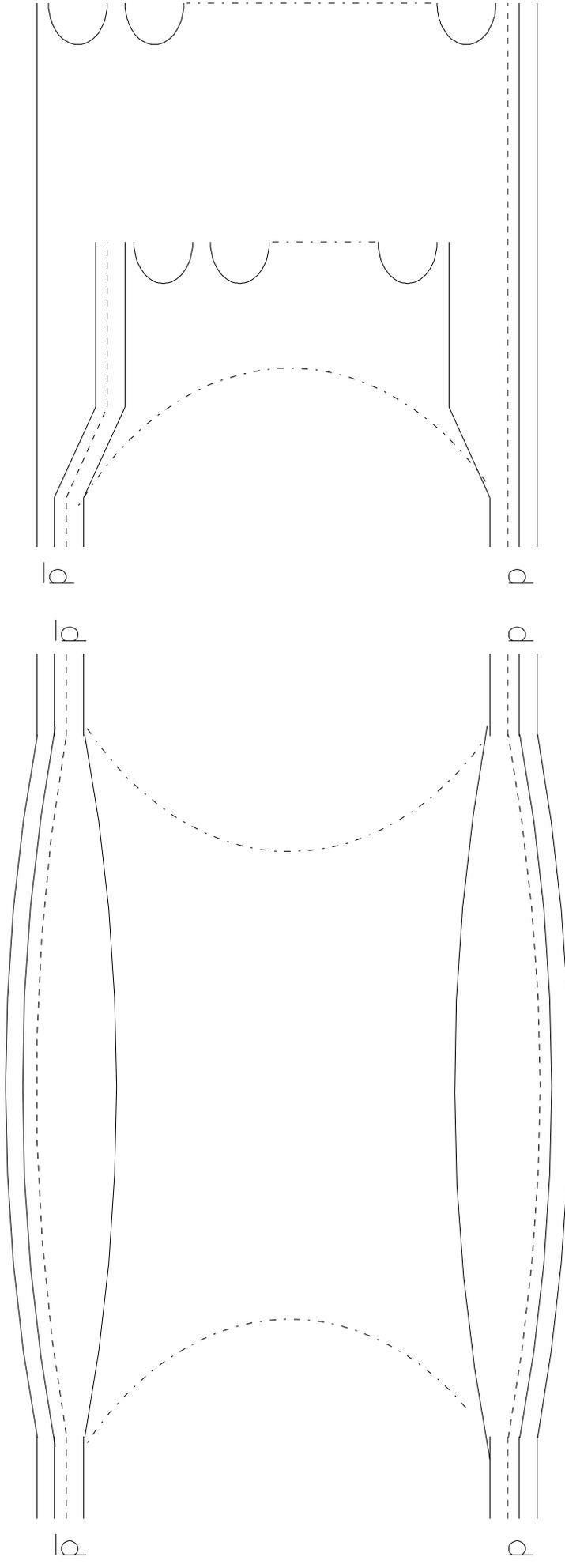
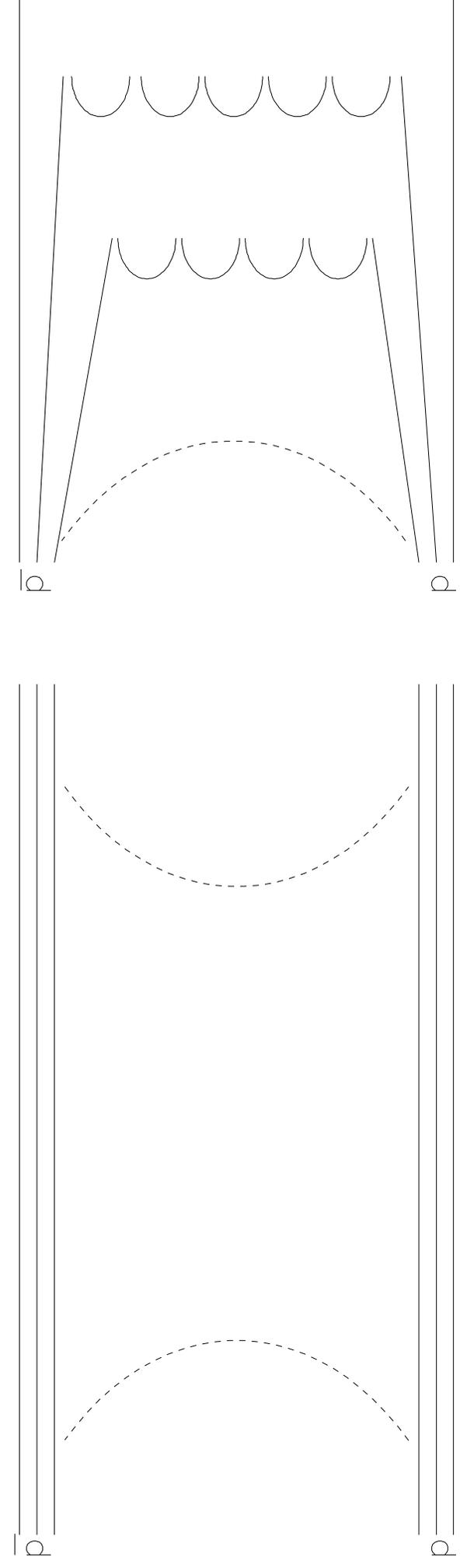
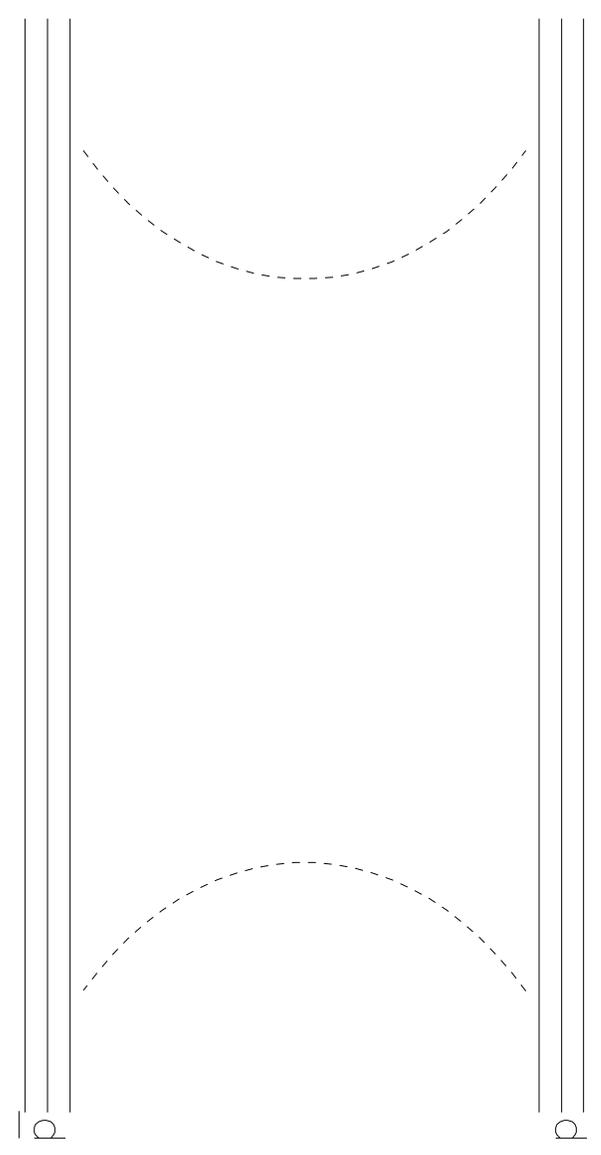
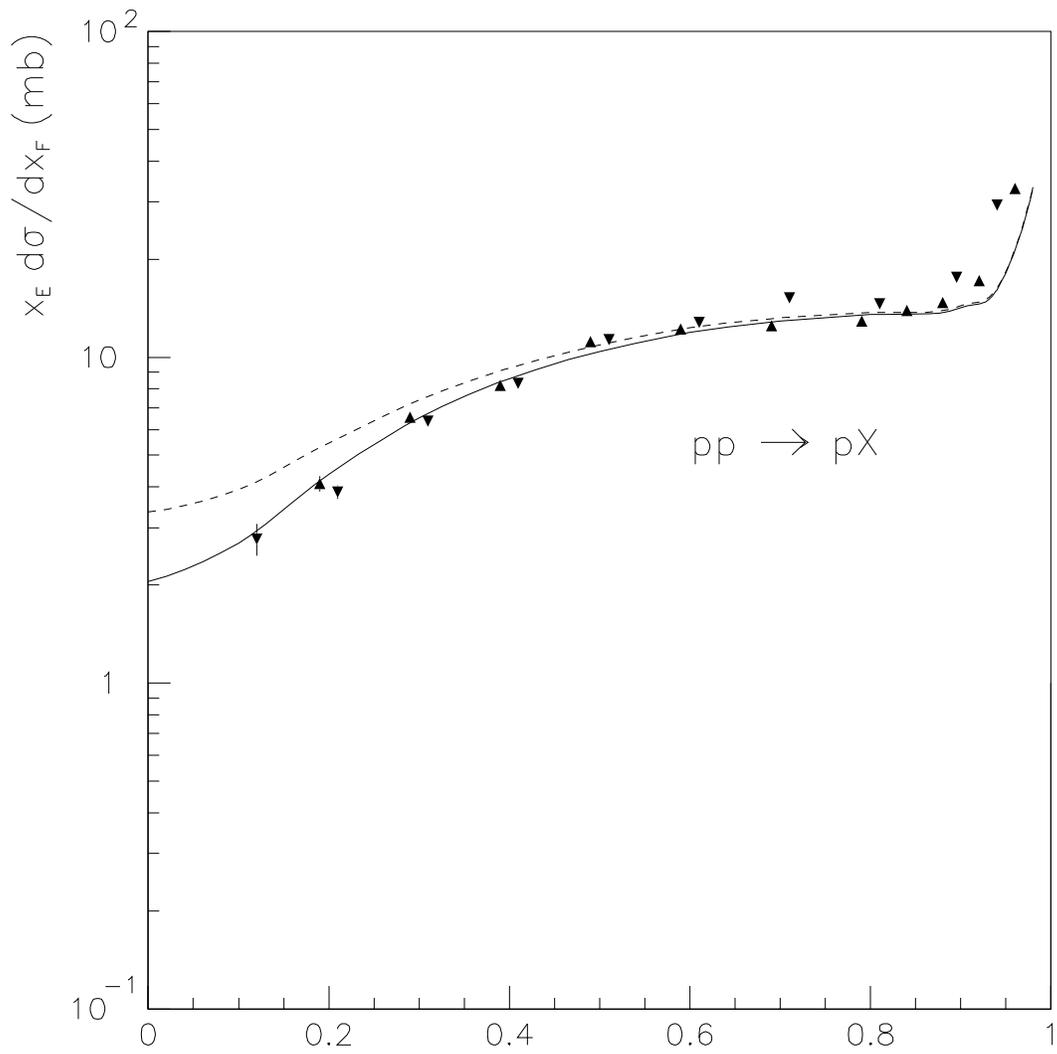
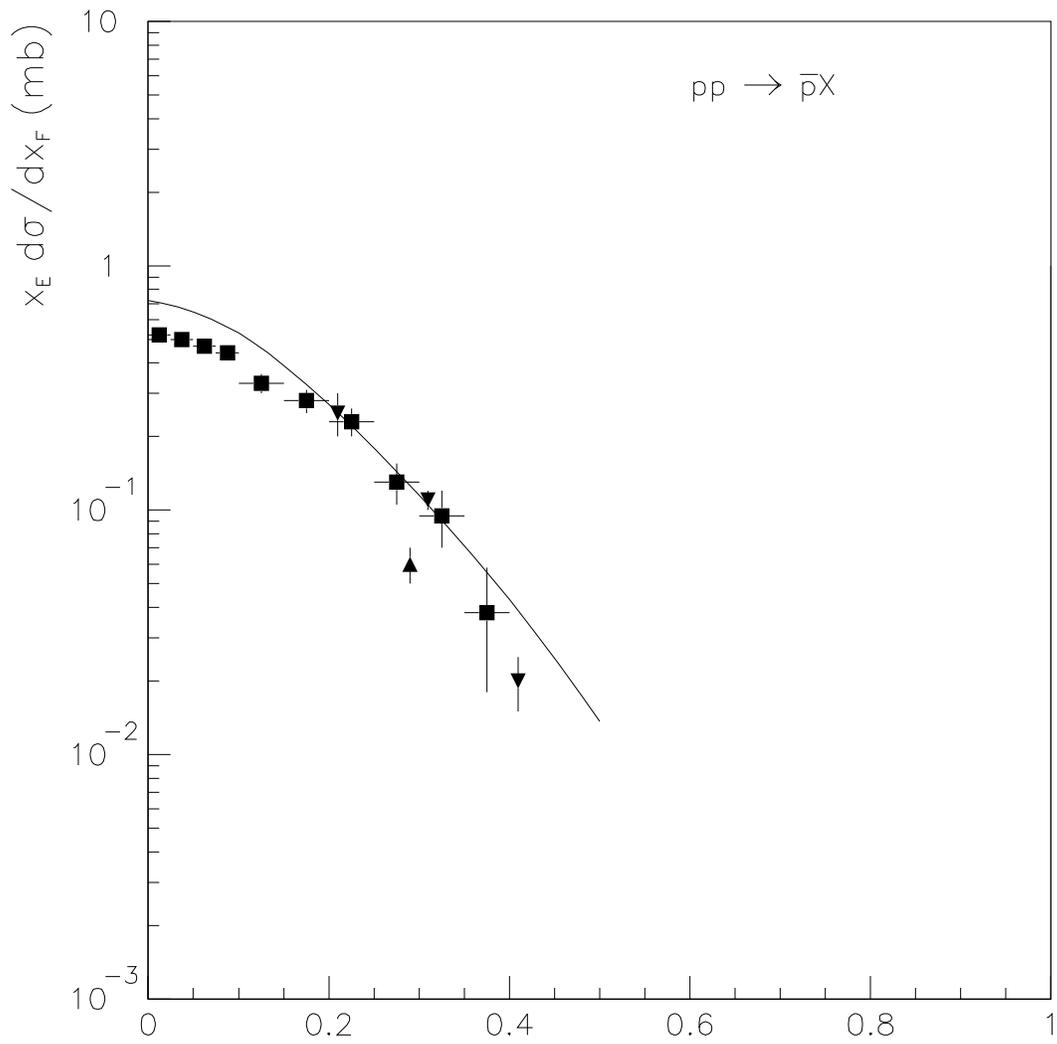
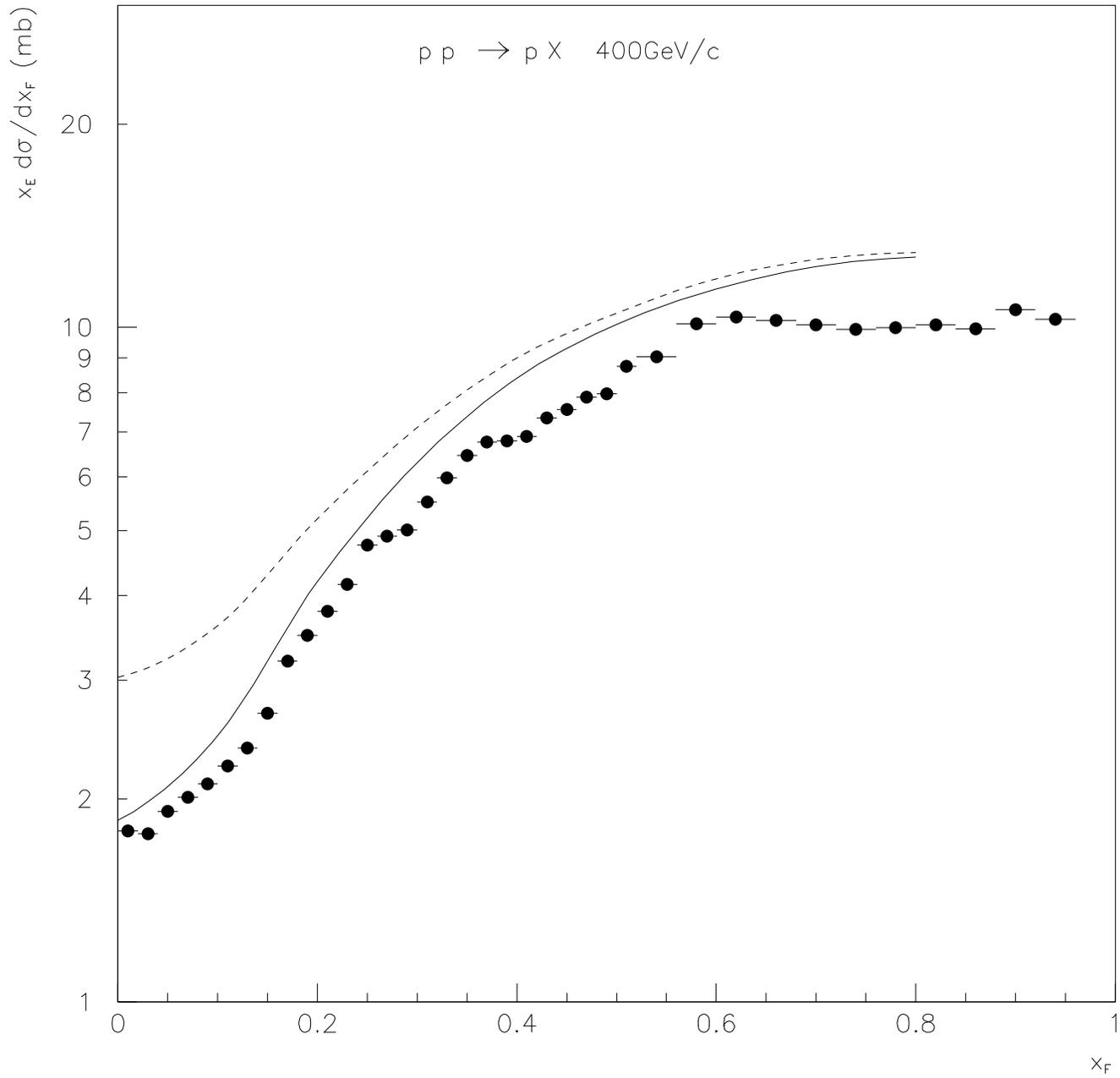
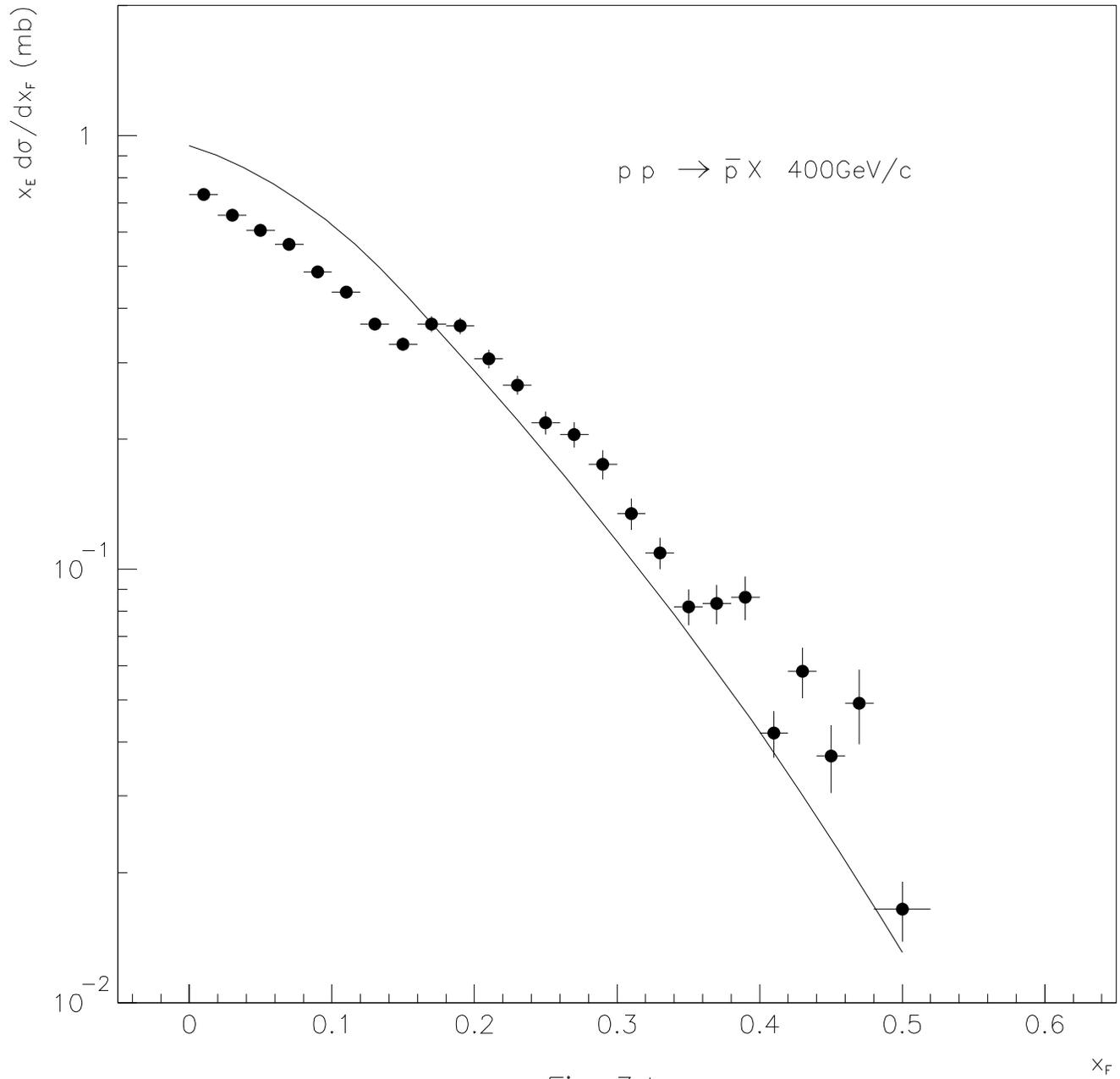
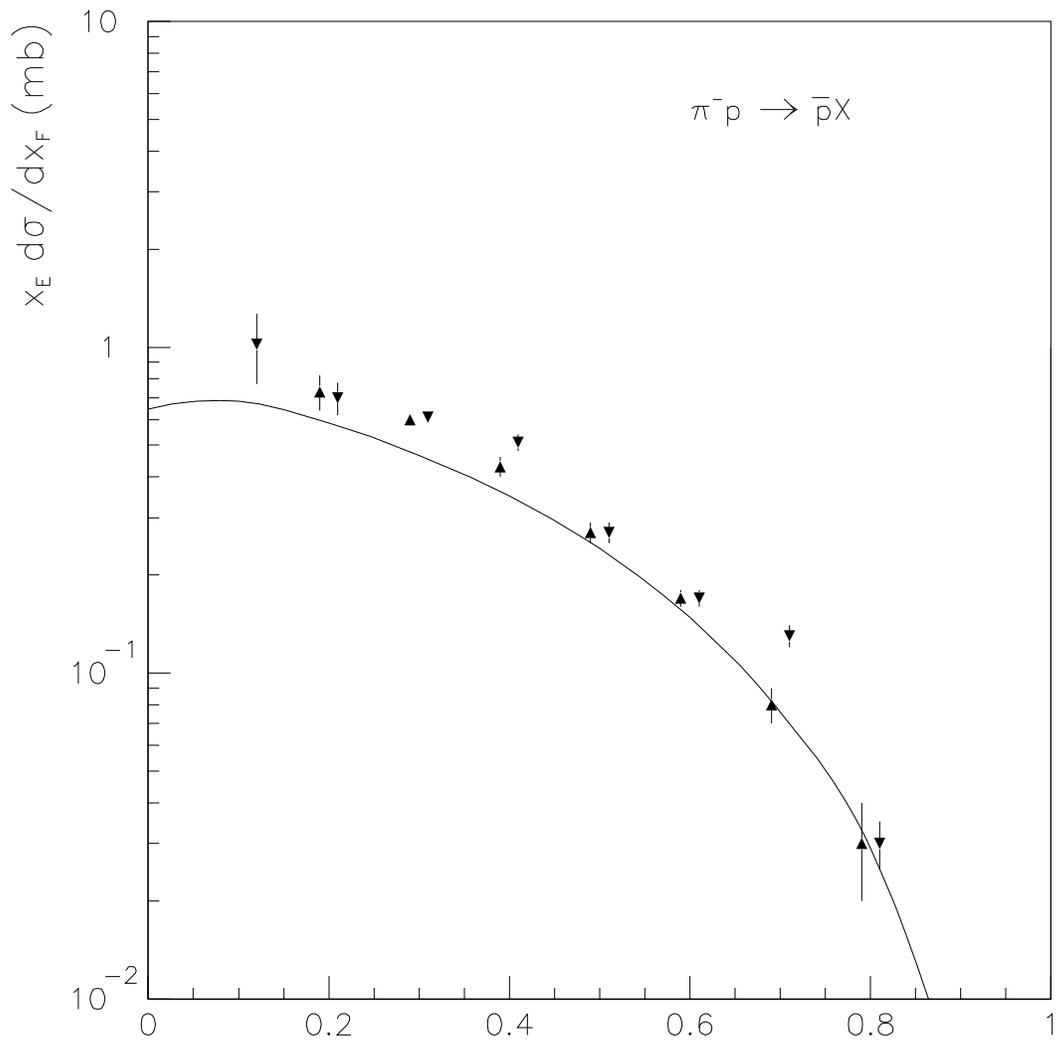
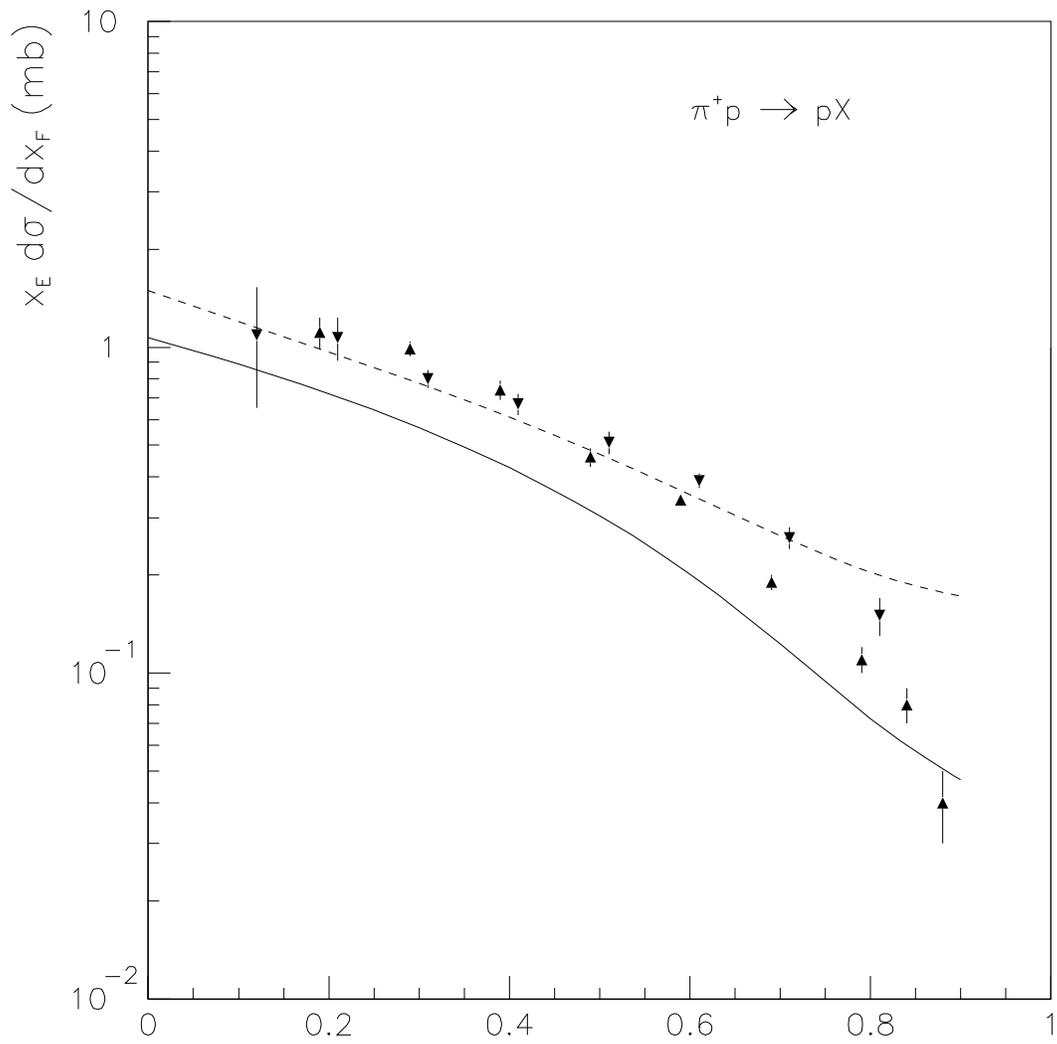
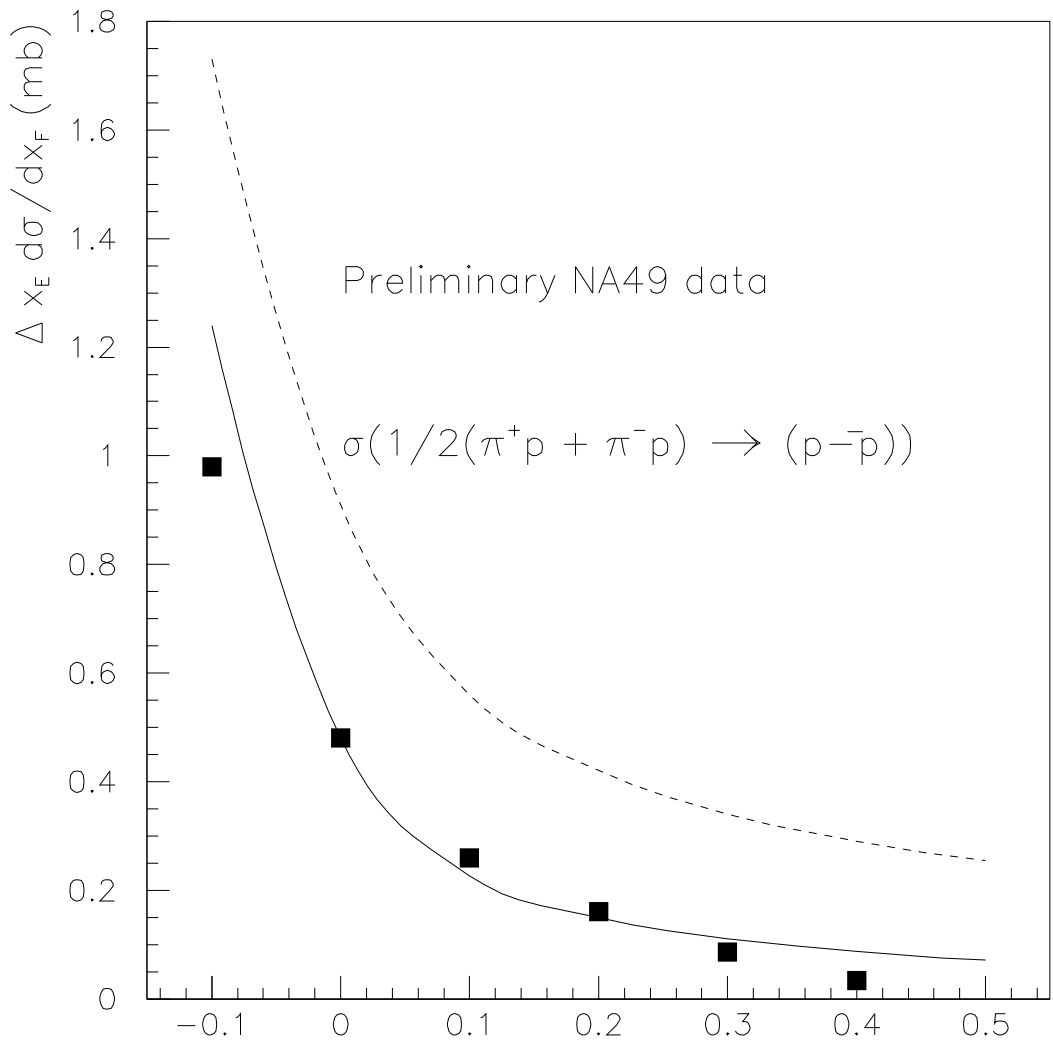
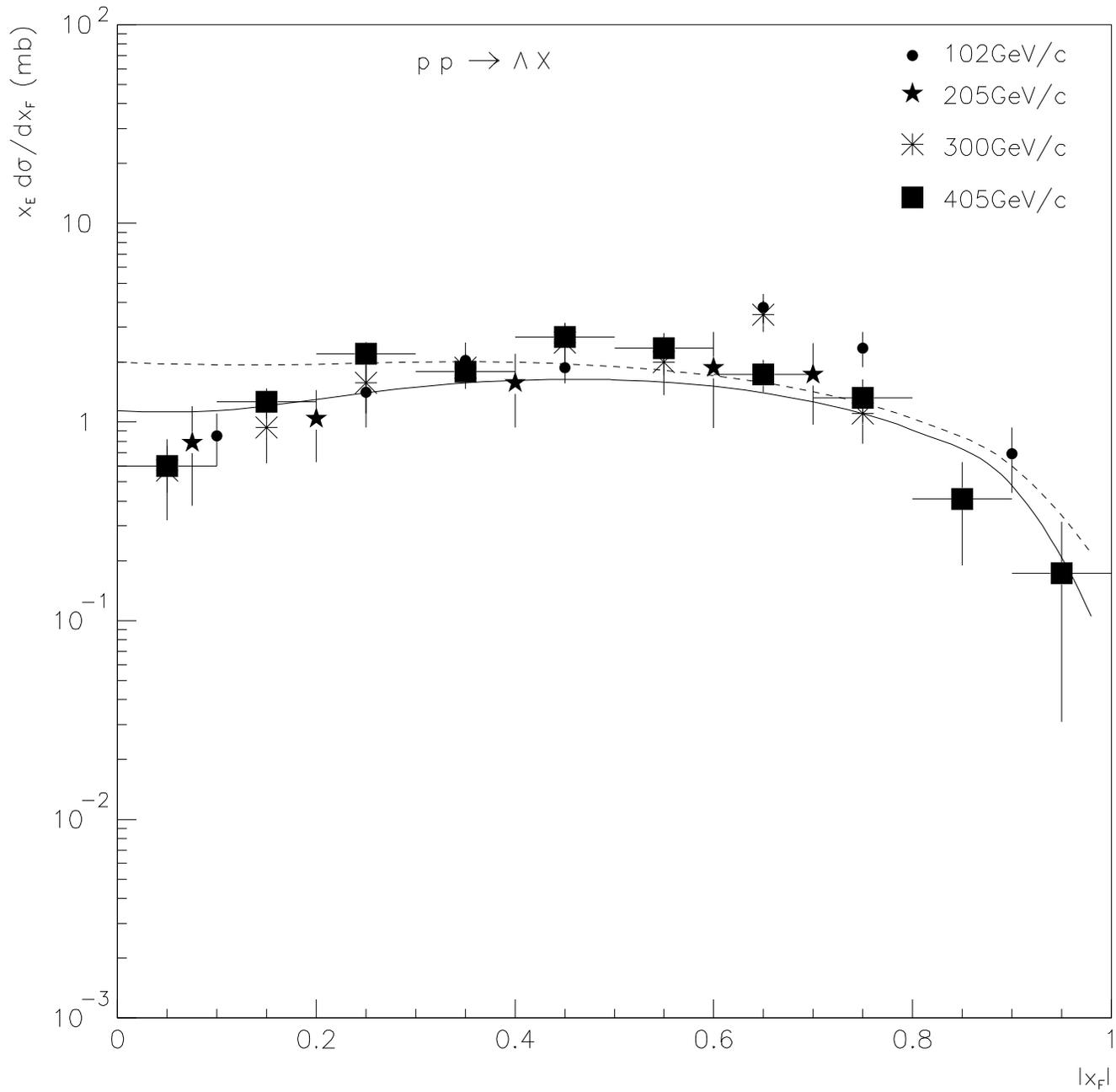
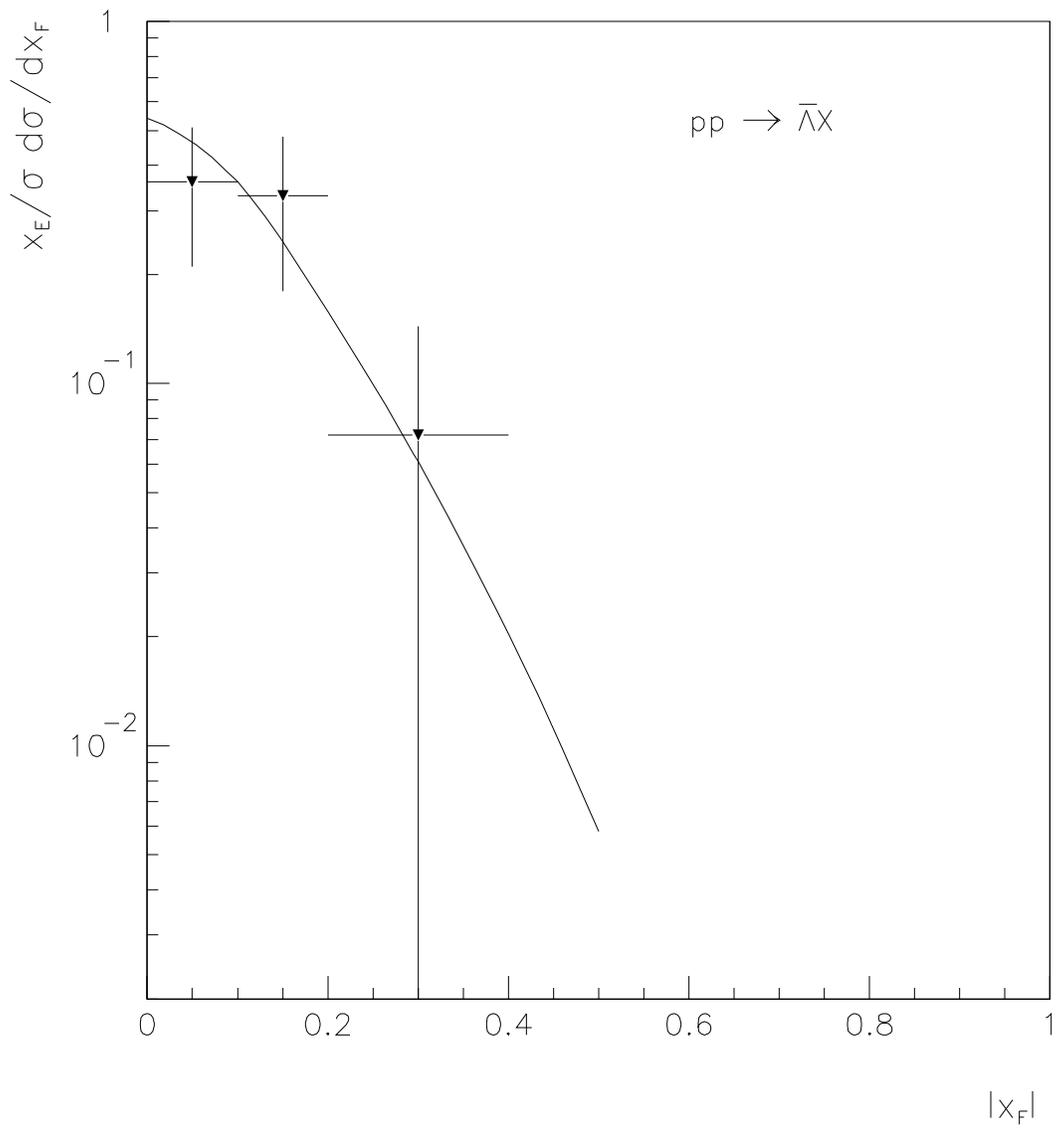
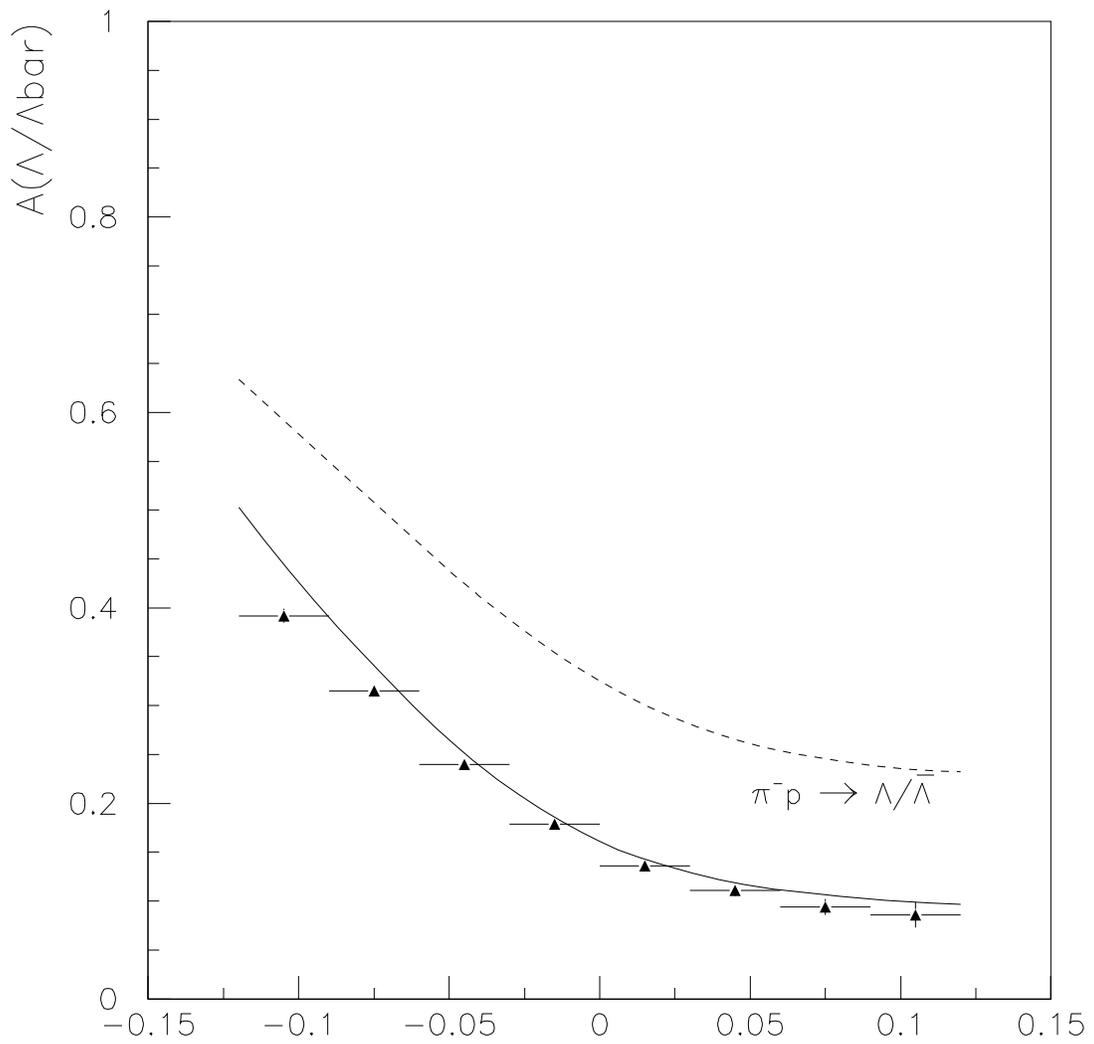
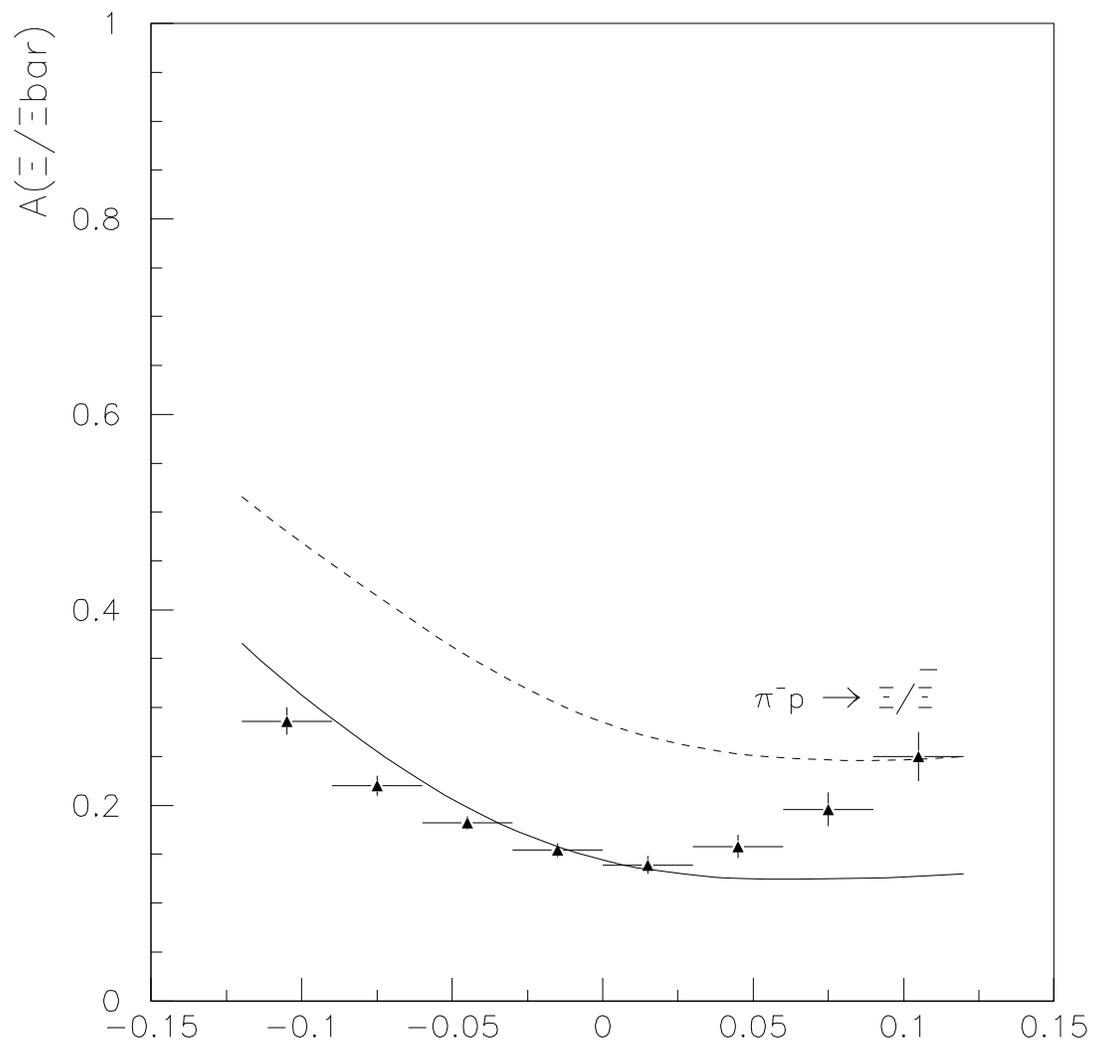
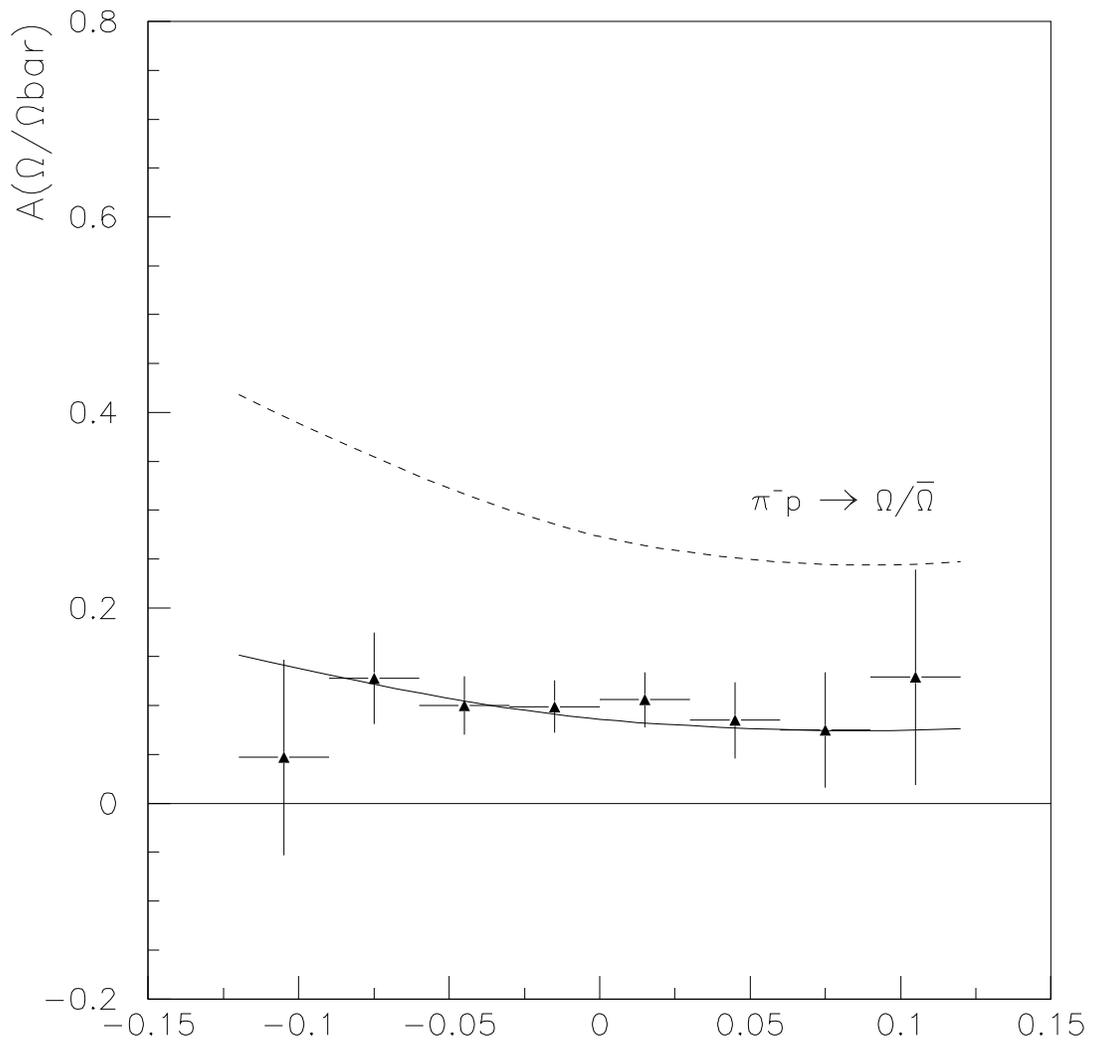
\end{document}